# Ultrafast laser high-aspect-ratio extreme nanostructuring of glass beyond λ/100

G. Zhang[1] , A. Rudenko[2] , R. Stoian[2*] , and G. Cheng[1*]

[1] School of Artiicial Intelligence, Optics and Electonics, Northwestern Polytechnical University, Xi'an 710072, China

[2] Laboratoire Hubert Curien, UMR 5516 CNRS, Université Jean Monnet, 42000 Saint Etienne, France

*Address correspondence to: razvan.stoian@univ-st-etienne.fr, guanghuacheng@nwpu.edu.cn

April 10, 2025

**Abstract**

The ultimate feature size is key in ultrafast laser material processing. A capacity to signiicantly exceed optical limits and to structure below 100nm is essential to advance ultrafast processing into the field of metamaterials. Such achievement requires to combine the control of optical near-fields and of material reactions, while preserving the exibility of long working distances, compatible with a mature laser process. Using sub-ps and ps non-diffractive Bessel beams, we demonstrate unprecedented feature sizes below a hundredth of the incident 1μm wavelength over an extended focus depth of tens of μm. Record features sizes, down to 7nm, result from self-generated near-field light components initiated by cavities induced by far-field radiation in a back-surface illumination geometry. This sustains the generation of more confined near-field evanescent components along the laser scan with nm pitch, perpendicular to the incident field direction, driving by local thermal ablation a super-resolved laser structuring process. The near-field pattern is replicated with high robustness, advancing towards a 10nm nanoscribing tool with a μm-sized laser pen. The process is controllable by the field orientation. The non-diffractive irradiation develops evanescent fields over the focusing length, resulting in a high aspect ratio trenching with nm section and μm depth. Higher energy doses trigger the self- organization of quasi-periodic patterns seeded by spatially modulated scattering, similarly to optical modelocking. A predictive multipulse simulation method validates the far-field-induced near-field electromagnetic scenario of void nanochannel growth and replication, indicating the processing range and resolution on the surface and in the depth.

## 1 Introduction

The accuracy and the minimal feature size in laser material processing are often associated with the tightest focusing of ultrafast laser beams, on the scale of the wavelength. How small the processed features can be and what are the limits of an optical process? The question of processing resolution limits is now of stringent actuality [1], determining a major development pathway in laser processing. The challenge of a laser process is then to exceed optical diffraction limits and to advance the laser structuring capability well below the processing wavelength, thus providing processing super-resolution. Such a capacity becomes key for the challenges of a laser- based design of metasurfaces and metamaterials, i.e. a next level challenge in laser processing, while preserving the fabrication exibility and light accessibility on materials compatible with an application-driven laser process. The perspective is the nm level. Ultrafast processing has an intrinsic capacity to localize energy within the optical limits by minimizing diffusive energy transport. To tackle the optical limits, a direct ablative process may rely on the nonlinearity of excitation[2], a possibility that was recently challenged [3], stating that the threshold deines the resolution. If the ingerprint of a surface ablation process is threshold-related, the depth will nonetheless follow the nonlinear absorption proile. Both characteristics can be used to narrow down the spatial extent of an ablative ingerprint. Two elements of advancement were early recognized, firstly in the capacity of a tightly-focused laser beam to generate local nanostruc- tures in the 100nm range on surfaces and below, based on threshold dependence [4, 5], boosted in particular cases by pulse time engineering [6]. Secondly, potential was seen in the possibility to generate near-field and to trigger nanoablation regimes [7] using extrinsic near-field genera- tors in the form of tips or apertures. The capability to generate near-fields with their spatial spectrum uniltered by diffraction was considered key for the laser fabrication of sub-wavelength features in a surface process limited in depth the by skin-depth absorption. Recently, the two aspects merged in a single irradiation technique that combines tight-focusing and generation of near-fields by self-induced scattering features; i.e. scatterers induced by the irradiation light itself. Such techniques were capable of generating a variety of sub 100nm features on surfaces and in the bulk [8-11], approaching characteristic sizes of few tens of nm or below, and thus bypassing a milestone in laser processing. These include scribing in the 10nm range via con- inement control as reported in surface tight-focusing geometries [10, 12]. Given the role of near-field scattering and its characteristic spatial pattern, the resulting shapes and geometries are controllable by incoming laser polarization, suggesting shapes relevant for photonics applications, data storage, or stealth dicing with extreme resolution. The intrinsic concentration of near-fields and interferential nature of multiple scattering is equally playing the major role in 2D and 3D self-organization [13, 14], with accessibility to the smallest scales.

The process provides then a kick-start of a nm-scale optical near-field, with its range defined by the shape and size of the scattering center, equally induced by the incoming processing laser beam. Such center, of sub-wavelength size, may not necessarily have an ablative nature; rapidly quenched nanocavitation developing at the speed of sound will end up at sub 100 nm diameters [15, 16]. This later process requires strong light gradients, usually achieved by tight focused beams or strong nonlinearities, but may develop at surfaces and in the bulk  To comply with the requirements of   exibility, such combination is optimally triggered by incoming far-field light. Therefore non-diffractive beams, formed by conical intersection of waveforms, are natural candidates to extreme laser nanostructuring, proving long working ranges and tolerance to the geometrical complexity of the piece. Secondly, a non-diffractive beam can in principle develop high aspect processing in a single pulse [17], extrapolating the concept of near-field processing to the bulk. Using long range ultrafast non-diffractive zero-order Bessel beams we demonstrate unprecedented feature sizes at almost a hundredth of the incident 1um wavelength with high aspect ratio, responding thus to a major challenge in efficient processing, maintaining resolution on an extended range. This results from self-generated near-field light components supported by cavities induced by farf-ield non-diffractive laser interaction all along the non-diffractive length. Using a back surface irradiation of silica glass, the u-scale energy gradient imposed by the Bessel beam induces a cavitation process ending in a scattering hole of less than 100nm size. This sustains continuous near-field generation along the laser scan and propagation axis. Local material removal occurs, at surface and in the bulk. Upon scanning the near-field pattern is replicated with high robustness on surfaces and in the bulk, driving a thus super-resolved high aspect laser structuring process, advancing towards a 10nm nanoscribing tool with a u-sized laser pen.

We report here the generation of extreme sub-10nm feature sizes, to our knowledge, the shortest   reported so far with high aspect ratios exceeding 1000, using loosely focused non- diffractive ultrafast infrared beams in fused silica glass. We discuss the nature of the trigger as far-field induced cavitation and the evolution process as interface near-field nanoablation developing at surface and in depth forming of nm broad trenches, driven by the field orientation during the scan. We illuminate the interplay between the development of near-field components with an extended range in depth and the material dynamic evolution from pulse to pulse using a self-consistent multiphysical modeling, combining the electromagnetic and material removal processes.

The modeling approach captures the multipulse dynamics of nanovoids evolution in the scan and the associated processing mechanisms. We show that such features constitute a primary step towards a field seeded self-organization of more complex quasi-periodic patterns, controllable by pulse polarization and its time and energy characteristics.

## 2 Methods
### 2.1 Experimental setup

The back surface of a fused silica plate (Corning 7890) with a thickness of 3mm is illuminated with an ultrafast non-diffractive zero order Bessel Gauss (hereby called Bessel) beam of conical half-angle $\theta_{air}$ = 17° in air ($\theta_{glass}$ = 12° in glass). Such beams are generated by the conical interaction and interference of waveforms refracted by a conical phase. An ampliied ultrafast laser systems (Pharos, Light Conversion) is employed. The laser beam is mainly operated at 1030nm, with the harmonics being equally available (515nm) and used in speciic cases, explicitly mentioned in the text. The infrared non-diffractive length is 600μm in air (875μm in glass) and the beam diameter (full width half maximum) is $\Phi$= 2.405/ksin($\theta$) = 1.28μm (the beam diameter will be accordingly reduced for other wavelengths). The Bessel Gauss beam is generated from an axicon (apex 176°) and then imaged into the sample using a 4f imaging system with a demagnification of 20X. The beam crosses the sample and is positioned with the intensity peak located on the rear surface. The rear surface processing using a Bessel beam has the advantage of filtering the physical effects of the side rings, as the higher order rings, resulting of a higher order interference order, require during the crossing of the bulk the passage through a region with time-varying plasma densities [18], creating reflections and phase instabilities that smear the lobe profiles. Thus, the characteristic multiring profile is not imprinted on the rear surface. The laser repetition rate can augment to 400kHz and the pulse duration is tunable between 0.2 and 5ps. The scan speed of the beam relative to the sample is in the range 1mm/s, equivalent to 2.5nm pitch between two pulses at the highest rate. After irradiation, the samples were analyzed by various techniques of electron microscopy.

### 2.2 Model

Model and material parameters for fused silica are adopted from [19]. A Finite-Difference Time-Domain approach is applied to solve the system of Maxwell equations in three dimensions in the vicinity of the scatterer, ensuring theoretical above-threshold conditions, slightly diferent from the experiment. For each pulse and geometry, the nonlinear Maxwell equations were coupled with the rate equation for free carrier generation, electron and ion temperature dynamics (two- temperature model). The temperature dynamics is resolved for 10 ps until thermal equilibrium is reached and the maximum ion temperatures are established. The input optical source is Bessel shaped laser pulse with θ = 12° in glass, the peak fluence 0.3 J/cm$^2$ , pulse duration of 100fs and central laser wavelength of 1030nm. The pulse duration is considered generic, and the use of other pulse durations will only qualitatively influence the thermal levels, without affecting the overall scenario. Simulation implies the initial void channel with a 100nm-diameter and a 5 m depth inscribed on the rear side of the glass surface centered at [X,Y,Z] = (0, 0, -3) μm. The ultrashort laser source excitation is placed at the distance of 7μm from the rear surface, focused on the void structure. Scan speed is 12mm/s for both perpendicular and parallel scanning, corresponding to 30nm displacement of laser beam by the succession of single pulses. Due to extremely large dimensions of the problem and a 10nm-ine resolution required to model near-field effects, we limit our analysis to glass temperature distribution and define the regions with the maximum temperatures exceeding $T_{th}$ = 3×10$^3$ K that are likely to be decomposed and transformed into void-like structure after material relaxation. Such a temperature threshold accounts for a feedback driven by near- and far-field ablation effects, which are crucial in our study. The hot spots are further converted into newly generated voids that serve as the new initial geometric conditions for the next pulse simulation, considering also a spatial shift of the laser beam position perpendicular or parallel to laser polarization. We further simplify the problem by considering a low number of applied pulses (up to N = 30) and higher scanning speed than reported in the experiments. This is done to consider the far-field effects that occur while scanning over the distances comparable to laser wavelength in medium from the initial void nanochannel, which otherwise would require simulating several hundreds of pulses with experimental scanning speed. Furthermore, we do not consider any multipulse incubation processes, except for pulse-by-pulse void growth, and neglect heat or stress accumulation between the successive pulses.

## 3   Results and discussion

The capability to create a scatterer from sub-100nm holes and to generate photowritten lines close to 10nm width with an ultrashort non-diffractive beam is generically illustrated in Figure 1 in the case of a silica glass material. Fig.1(a) shows a typical hollow channel ending (section) induced in the material by a single non-diffractive laser pulse close to the processing threshold (in the range of $1J/cm^2$ ). Such a void-like feature is associated with a process of volume cavitation, as long as the local energy density exceeds the specific cohesion energy of the material [20]. The process is triggered by high energy concentration and strong gradients, either by using high cone angle beams or temporally stretched light pulses, avoiding thus plasma defocusing and deleterious energy spread [16]. Such a dependence justifies the temporal envelope control employed here. It has been indicated that the formation of such a nanochannel can be linked to the prior softening or melting of the material [16]. The nanosized void is then induced by the stress-assisted relaxation of the laser-induced liquid column, preserving the beam geometry, with a cavitation process being triggered in the liquid phase. Such a process is energetically favored as cavitation in a liquid phase has low mechanical requirements and the conditions are easily met. A resulting void-like structure with a high index contrast can scatter light. The near-field components generated by the scattering hollow channel are also depicted in Fig.1(a). The scattering pattern has two main components; a surface pattern and a volume pattern of similar energy but embedded in the material. The pattern presents, perpendicular to the optical field, lobes in excess of 50% field enhancement, above the processing threshold, and a reduction of intensity sidewise, arresting the lateral evolution of the process. Such a pattern will be multiplied in a scan and further sustained as the scattering feature acquires a more elliptical shape. The result of a single line scanning in the conditions of Fig.1(a) is shown in Fig.1(b), achieving a trench diameter of 15nm. The figure illustrates thus the current ansatz of the process, the hypothesis of a far-field radiation generating ablative near-field components via a self-induced scatterer. Similar situations based on tight focused Gauss beams were recently reported for surface and volume nanoprocessing with a characteristic size well below 100nm [10, 11]. The present situation shows that a extreme nanoprocessing scales can be obtained by rather loosely converging radiation, with a moderate cone angle and a spot size exceeding the micrometer. We will discuss below the enabling factors and the physical mechanisms behind.

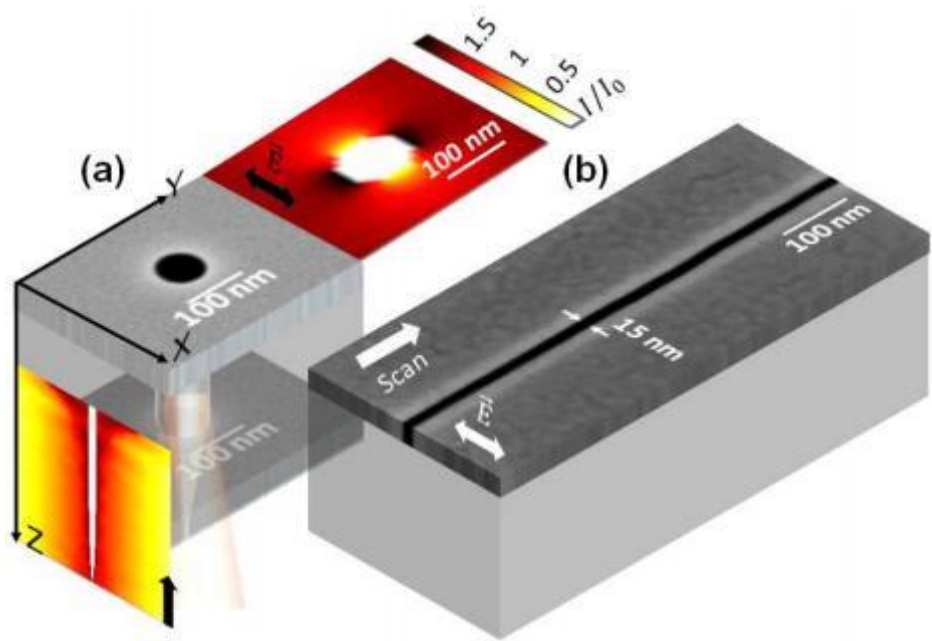

Figure 1: (a) Typical nanohole (top of a channel) generated in bulk fused silica by single short pulse non-diffractive Bessel-Gauss beams of moderate cone angle. These void channel structures can extend to the rear surface. Such structures can be obtained for a variety of incident conditions of cone angle, pulse duration and illumination wavelength. (for example for infrared (1030nm) and visible (515nm, present case) radiation at μJ energies). The illumination of such structure generates near-field evanescent components with the poles oriented perpendicular to the laser field. The color map indicates the intensity enhancement relative to the incident value for an infrared beam scattered by a void channel of similar geometry. One notices the field level increase perpendicular to the incident optical ield. These waves are perpetuated in the bulk, all along the channel, with a distribution of intensities given in the insert. (b) Photoscribed nanoline by linear scan of the laser with a section of 15nm. A 2ps pulse duration at 1030nm was used at 333kHz and 1.2mm/s speed scan.

### 3.1 Generation of high aspect ratio nanoscribing

Let us follow now the genesis of photoscribed lines at $\lambda/100$ feature size. The first element is the generation of a sub-wavelength scatterer. The cavitation process where a cavity forms due to energy gradients and evolves at maximum the speed of sound, being quenched by sub-ns diffusive energy loss, is the main process behind the achievement of a small void scatterer with a maximum index contrast. This initial scatterer at 1μm incident wavelength has a typical diameter in the sub 100nm range. This can be reduced by the employment of shorter wavelengths, suggesting perspectives for even smaller scales. This type of behavior is illustrated in Fig.2(a), showing the effect of energy on the size of a hole induced by a single 3ps at 515nm wavelength crossing the bulk and exiting the rear surface. Close to the threshold, a record small diameter in the range of 10nm was obtained, stressing the potential of nanocavitation for extreme nanoscale features. This is to our knowledge, the smallest feature induced by a static, direct focusing of an ultrashort laser pulse in the visible range, which shows prospect for breaking the 10nm milestone at smaller wavelengths. This element is the kick start element for nanoscribing, with a scan direction perpendicular to the field direction. Consequently, a narrow line is formed, as illustrated in Fig.2(b). Thus, triggered by a material process generated by far-field, the nanoscribing is then sustained by scattered near-field evanescent waves. An insightful description of near-field scattering on a laser-induced nanosized scattering center, perpetuating the size upon scanning was given in [11]. A similar scenario is at work here, where the non-diffractive beam characteristics will be reflected in both surface and volume near-field components generated by incoming pulses. The consequences of such distribution are manyfold during the scan. It is to be mentioned the combination of two effects, mechanical and optical; the material cavitation on the sub 100nm scale and the field enhancement at the edge, which leads to the apparent super- resolution in the processing size. The approach harvests thus the synergetic action between material response under light gradients and the induced light enhancement and confinement. At no time, the diffraction limit for the incoming far-field radiation is violated.

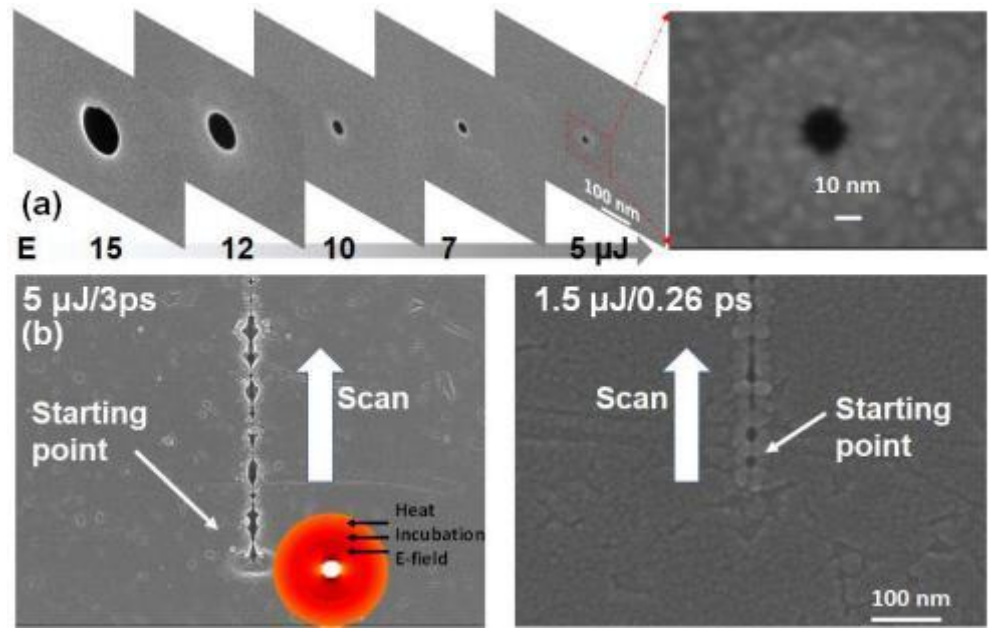

Figure 2: (a) Size control of Bessel induced nanoholes (channel sections) as a function of the beam energy. A minimum hole diameter close to 10nm (15nm) is obtained at the lowest energy, illus- trating the extreme scales due to nanocavitation. The processed channels were obtained with single pulse Bessel-Gauss beams at 515nm incident wavelength and 3ps pulse duration. Geometrical conditions: $\theta_{glass} = 11°$. (b) Such a nanohole serve as triggering element initiating the nanoscribing, as illustrated in the image for two different processing conditions (5 μJ at 3 ps and 1.5μJ at 0.26ps for a scan speed of 1.2mm/s at 333kHz repetition rate). The orientation of the surface near-field is given in the inset, together with an illustration of cumulative effects.

First, it is remarkable to notice two regimes of scribing, one showing smooth lines of less than 10nm section and, at similar energy input, a stronger regime of interaction, with a larger diameter and chipping of the edges, as can be seen in Fig.3(a). This bistable behavior can conduct thus to two stationary processes. Dimensionally, both regimes are self-sustained as the respective section stays constant, being solely defined by the size of the scatterer and the intensity of the scattered field. The transition from a domain to the other may be generated by an instability in irradiation or local threshold, which will locally modify the thermodynamic evolution of the material. Concerning the aspect, such a behavior can be associated with a change of material parameters in terms of ductility during processing. The present conditions, with a scan velocity of 1.2mm/s for an input repetition rate of 333kHz indicate a pitch of 3.6nm. Can such a small pitch at a high repetition rate be the source of a cumulative process? The input size (FWHM) of the beam is 1.28μm, indicating an accumulation of multiple pulses across the scan. Given the effective ablative area of few nm, the number reduces drastically to few pulses on the processed region, still on a background of fluence distribution in the vicinity of the threshold. Considering a diffusivity for fused silica of $D = 0.9mm^2/s$ and the characteristic time of diffusion $\tau_{diff} = w^2/D$, with w being the transverse size of the source, the heat transport will be the superposition of two sources, one, nm in size, with a cooling or diffusion time in the ns range, the other, μm in size (laser spot), with a cooling time in the μs range. In principle, at the present repetition rate, heat accumulation is possible without producing ablation effects, but it may, presumably, generate viscosity variations, explaining expectable changes in the brittleness of the glass. A second process, based this time on defect incubation will also lower the threshold around the processing spot, favoring thus more light localization. The process is based on structural distortions induced by trapping charge on molecular orbitals, transient in nature [21], and resulting in permanent broken bonds and the onset of color centers. They have a stronger optical coupling to the light in the incubation area than in the pristine matrix but they also affect the mechanical stability of the matrix. Thus, a nonlinear interplay emerges between heat and other structural cumulative processes, with their action range being illustrated in the inset in Fig.2(b). A second observation should be made. The typical defect rate generation in a pulse lies in the range of $10^{18}$-$10^{19}$ cm$^{-3}$ , a fraction of the excited carrier density [21]. That implies that, in the volumes of concern here, with nm ablation range on unit of length in the depth and nm pitch, the presence of defects in those region is scarce, which increases in principle the stochastic nature of the process and its nonlinearity. Nonetheless, the defect incubation however stops in the processing area as here, in the material removal range, the local temperature in the area of field confinement exceeds the softening temperature, putting a halt to the defect generation and stabilizing the process. Thus the transition between the two regimes can be triggered by small random instabilities in the input beam or material state that will be nonlinearily amplified due to the positive feedback between the strength of the field, the size of the scatterer and the local modification of the processing threshold. The process will be arrested by the increase of the density of the carriers, which, by their anti-phased oscillations in the field, will scatter and delocalize radiation, decreasing the energy concentration. Similarly, as seen, a stabilizing effect is expected from the temperature. This mix of positive and negative feedback regulation confers robustness to the scribing process. This robustness is seen in the uniformity of the process and in its unaltered depth range.

The second relevant question is if the process is a superficial one, located at the surface. Focused ion beam milling and inspection by electron microscopy indicated that the nanoscribing extends deep in the bulk, creating nm (10-20nm) section trenches of several tens of microns depth, achieving thus remarkable aspect ratios in excess of 1000. This situation, illustrated in Fig. 3(b) is characteristic to both ranges, gentle and strong scribing regimes. This design of the near-field on surface and deep in the bulk is intrinsically related to the non-diffractive nature of the irradiation and to the geometrical shape of the initial scatterer, a one dimensional void-like feature with conical geometry.

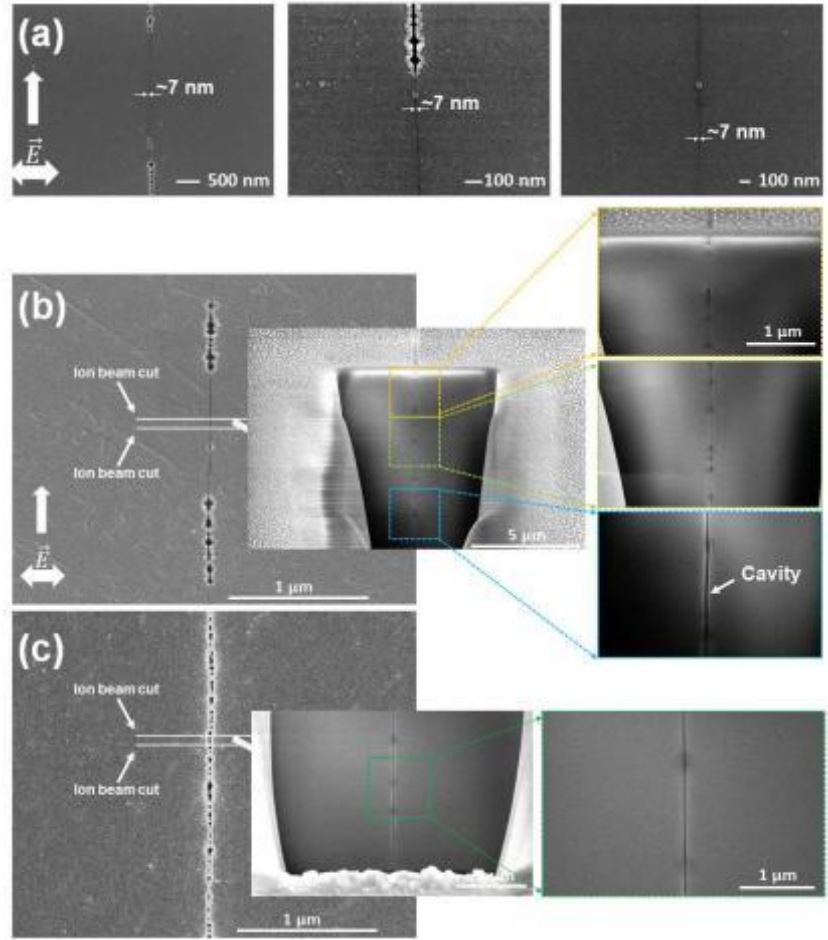

Figure 3: (a) Bistability in nanoscribing. The topography changes from chipped aspect along the scan to smooth and uniform line. Processing conditions: 1030nm incident wavelength, 3ps pulse duration, 333kHz laser repetition rate, 1.2mm/s scan speed. (b,c) The inspection of the depth of the nanoscribed channels indicates similar nanostructuring in the depth for both nanoscribing regimes.

### 3.2 Vectorial process control

A key issue is related to the controllability of such a process. Having established the role of near-field enhancement in driving the nanoscribing process debuting with a first self-induced initiating center, and the role of irradiation conditions in kick starting the process, process control implies chiefly the orientation of the near-field domain. A scattering process depends on the direction of the oscillating electric field and on the contrast of the dielectric function at the borders of the scattering domain. The challenge is to align the near-field concentration on the direction of the scan for a given shape of the scatterer and a given dielectric function of the excited or transformed material. Thus the direction of development of the process becomes critically dependent on the incident polarization. For the given material dielectric function, which remains dielectric under the present conditions in spite of electronic excitation, which stays sub but close to critical density [22], the scattering around a hollow channel tending to become elliptical, occurs in the glass environment perpendicular to the field. This offers a control knob for designing complex scribing trajectories. A circular trajectory is shown as example in Fig.4(a,b) providing scribing aspect on the surface and in the bulk. The similar development on the surface (Fig.4(a)) and in-depth (Fig.4(b)) demonstrates the robustness of the process.

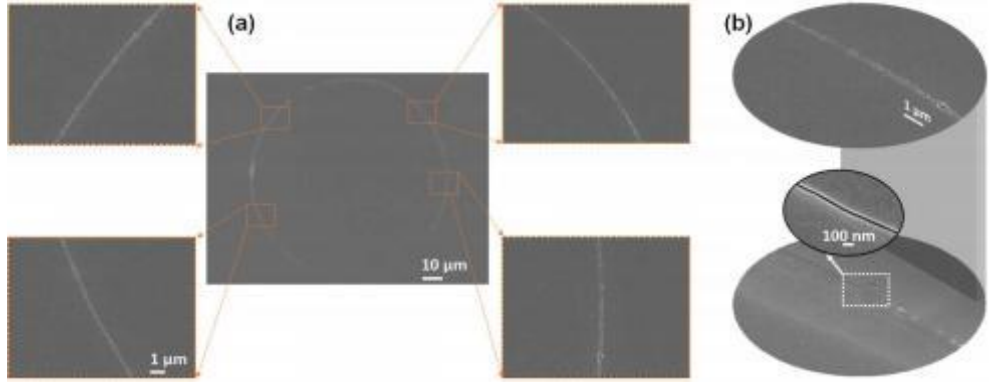

Figure 4: (a) Curved trajectory for nanoscribing in the strong range, with dynamic polarization control during scan, adapted (perpendicular) to the local curvature. The scribing conditions are: 3ps pulse duration, 5 μJ input energy, 1.2mm/s scan speed at 333kHz pulse repetition rate. (b) The scribing develop robustly in the bulk as revealed by ion milling of scribed surfaces in (a).

### 3.3 Modeling laser nanoscribing

A multiphysical modeling approach was applied to identify the multipulse dynamics of void evolution on these extreme spatial scales (section 2). It includes the description of the three- dimensional nonlinear propagation of the Bessel laser beam in the vicinity of the void nanochan- nel, the evaluation of the scattering patterns and their evanescent components, the corresponding local nonlinear excitation, and the estimations of the electron and ion temperature distributions both on the surface and in the bulk of the glass sample. Based on these temperature charts, the regions affected by near-field ablation are specified according to a thermal criterion corresponding to vaporization. The affected regions are further considered as newly void scattering structures for the next pulses. We analyze the nonlinear fluence distributions and temperatures, indicating how the energy is deposited, absorbed and converted into heat inside the material after excitation by multiple pulses, shifting the position of the laser source in the direction of scan for each considered laser pulse. As a result, the interplay between light and matter is investigated, where light modifies the material which in turn will modify the distribution of photons in a dynamic manner. The details of the model and its assumptions are discussed in the 2 section.

The mechanism of void nanotrench growth is described in Fig.5 based on the distribution and the level of the scattered energy. Fig.5(a,b) recalls the experimental situation, with the two nanoscale scribing regimes. Surface and volume proiles are given. Fig.5(c,d) shows the calcu- lated distribution of fluences F and ionic temperatures $T_i$ for an increasing number of pulses, taking into account a displacement of subsequent pulses during the scan, emphasizing both far- field and near-field scattered components. The conditions defining the size of the initial scatterer were chosen to allow reasonable sampling and to reduce the calculation time but the model is scalable to smaller sizes. The hybrid multipulse electromagnetic and ablative model confirms the achievement of the nanoscribing process along the scan when the field is perpendicular to the scan direction. The elongation of the processed area with increasing dose is becoming apparent. The scribing is an erosion-like process, only driven by the leading near-field lobe (Fig.5(c)), but not multiplying cavitation events. The level of the enhanced, near-field peak fluence is approaching $0.5 J/cm^2$, in augmentation of 10% with respect to the input peak fluence. The calculation of excitation conditions, namely of the absorbed energy shows that in the near-field regions, temperatures in excess of $3\times10^3$ K can be obtained (Fig.5(d)), sufficient to generate nanoablation on the cavity walls via a mix of evaporation and possibly nucleated vaporization of nm thickness material. The process becomes superficial, consistent with superficial ablative regimes on the surface and on the inner channel walls, at the glass interface, with local temperatures exceeding the standard vaporization ranges and carrier densities coming close to critical ranges for the 1030nm radiation. We consider vaporization concepts acceptable as threshold criteria for the small volumes involved, similar to the critical size of nucleation centers, even though the process is confined to few nm. It is of interest to see that the process is developing similarly in the bulk, all along the input beam propagation axis in the bulk (perpendicular to the rear surface). Fig.5(e) shows the near-field nonlinear fluence distribution around an initial channel (induced by the one-dimensional Bessel cavitation). During the scan, the cavitation process, responsible for the formation of the initial void nanochannel, does not multiply, the accumulation of Bessel lines near a first cavity does not further provide a sufficient energy to sustain cavitation. Alternatively, the near-field enhancement at the edge promotes the trench processing by an erosion-like process at the edge (as recalled experimentally in Fig.5(a,b)). It is interesting to note the very confined range of action of these local processes. This validates the fact that cavitation is not further supported during the scan and the process is solely driven by near-field erosion at the edges. Such a near-field abrasive process driven by temperature has a tendency to decrease the feature size with increasing the dose (see section Supplementary Material), i.e. with decreasing the pitch between two pulses to few nm. The process occurs at the tip of the near-field lobe. The spatial range of the near-field action becomes thus more confined, and smaller than the initial seed size. The conditions for further reduction of scale are thus met.

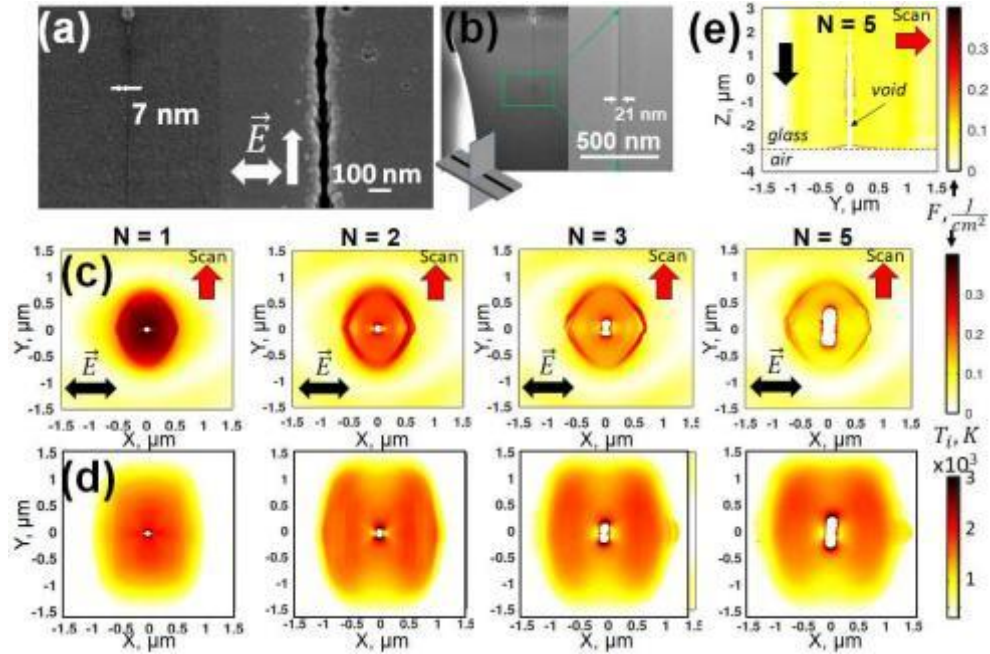

Figure 5: (a, b) Experimental images for nanoscribed void nanochannel (a) on the rear surface (illus- trating the nanoscribing ranges) and (b) in the volume (the depth proile corresponds to the strong scribing case). (c) Nonlinear fluence distribution F around the void nanostructure for an increasing number of applied pulses N on the rear surface Z = -3μm. (d) Ion temperature distributions showing localized increase in the temperature in the near-field regions. (e) Bulk development of near-field components and nonlinear fluence distribution in the volume at X = 0 (the position of the initial void channel at Y = 0 and the glass-air interface are indicated). Scan direction is perpendicular to laser polarization.

### 3.4 Towards self-organisation

Increasing the processing energy, one departs from single line scribing and patterns of a certain quasi-periodicity will appear, similar to previously observed laser-induced periodic surface and volume structures [13, 23]. Such situations are described for different scan directions relative to the polarization in Fig.6 and Fig.7. Inspecting the experimental situation in Fig.6(a), one observes the seed of new nanolines parallel to the main line appearing with increasing energy. The pulse duration affects the energy interval for observing the transition from single to double and to multiple lines, inluencing at the same time the spacing between the lines and thus, their quasi-periodicity. The process multiplies with increasing energy and new sidelines appear. The modeling results indicate that the generation of adjacent lines is seeded by lateral scattering from the main line. The local action of scattered waves is a concept that was also found in surface periodic structures, and mimics here a passive optical modelocking process [24], where energy injection generates and stabilizes the neighboring structures. The factors are jointly at work in generating additional damage; the increase of the exposure level and the incubation level. The near-field pattern indicated in Fig.1 presents, as we have discussed, enhancement lobes in the direction of the scan and depletion lobes lateral to this. Laterally the field recovers to the original background value within a sub-wavelength range, increasing the probability of damage when it stabilizes at the background values. Such a process was suggested in the context of nanogratings generation by Liang et al.[25]. Being related to the probability of damage, the process is aperiodic by nature, but self-organized under the action of light, where coherent scattering effects may render a certain periodicity in relation to the wavelength.

In the simulations, the threshold for a secondary void structure formation is reached when the initial void channel is enlarged into an elliptical shape by near-field effects ($N = 5$), and starts to scatter light stronger not only in the vicinity but also at distance. The interference between this scattered field and the incident light from a focused Bessel beam creates a standing wave pattern with stronger nonlinear fluence enhancement, higher electron density and temperature. Once the temperature threshold is achieved in the extrema points ($N = 10$), the affected areas are considered as new void seeds that enlarge by the near-fields upon the next pulse irradiation (see $N = 20$). Upon the growth of two neighbour void channels, the maximal field enhancement and temperatures are reached as well in between the channels at the upper positions intensified by scanning from bottom to the top, as indicated in Fig.6(b) for $N = 25$ applied laser pulses. This leads eventually to the formation of three aligned void nanochannels ($N = 30$), separated by sub-wavelength distance. The exact spacing is defined by several parameters, such as focusing conditions, scanning speed and laser pulse energy. The energy affects the growth rate of the void nanochannels, therefore, their ability to scatter light as they grow and the position of the field enhancement extrema in the far-field and in between the structures. The scanning speed and focusing conditions set the limit how long the area is affected by laser and the near- and far-field patterns are relevant, i.e. whether the void structures have sucient time to grow and to replicate. The process is, similar to the single line formation, driven by local temperature increase and nanoablation (Fig. 6(c)).

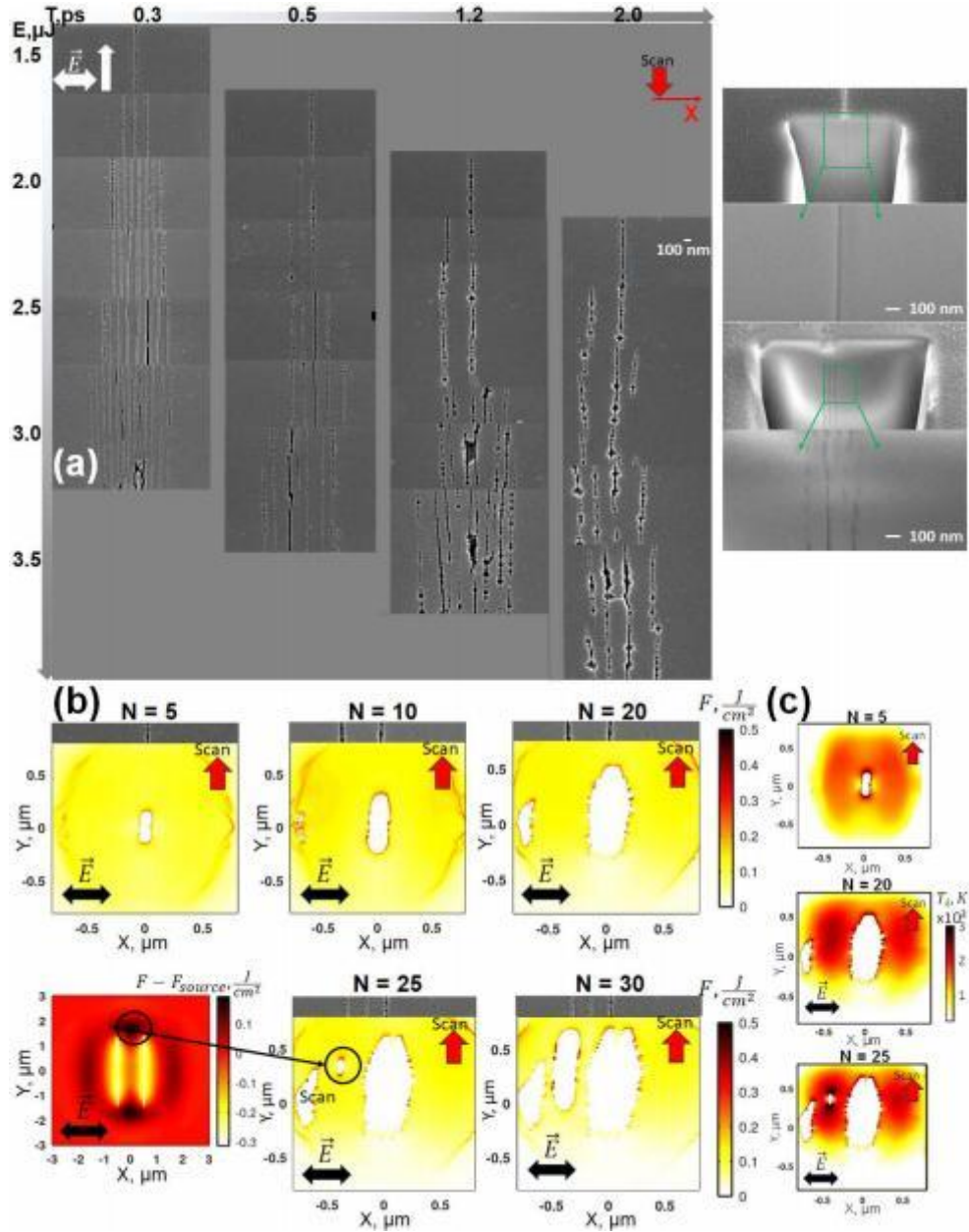

Figure 6: (a) Evolution of quasi-periodic traces with the pulse duration and energy, for a scan direction perpendicular to the laser polarization. The apparent periodicity and the efficiency of line multiplication depend on the irradiation conditions. (b) Simulation of additional damage lines as a combination of near-field recovery and far-field scattering overlapping with the incoming field. The sub-figures show the nonlinear fluence distribution F on the rear surface of glass at increasing number of pulses N while scanning perpendicular to the field polarization, starting from an initial 100 nm-nanovoid in the center of the grid [X,Y] = (0, 0). The appearance of a new void structure in between two void nanochannels at N = 25 is attributed to the linear fluence enhancement between two channel scattering sites underlined in the left inset. (c) The corresponding selective temperature charts, illustrating the conditions for nanoablation in the ield enhancement regions.

Similarly, a quasi-periodic pattern with energy-dependent periodicity emerges also in con-ditions of scan parallel to the polarization. A sequence of parallel structures is formed where individual features are elongated by the near-field directional erosion and side damage appear with a probability defined by the scattering pattern.

The experimental situation is given in Fig.7(a). The model illustrates the formation of secondary damage sites sustained by the su-perposition between far-field components and background energy density, already close to the damage threshold. Fig.7(b) shows first the deformation of the initial structure under multiple exposure, acquiring an elliptical shape. Side scattering reinforced by the scan, produces side damage lines that soon acquire an elliptical shape. The combination between coherent scattering effects and damage probability creates the quasi-periodicity character, with an intrinsic partially stochastic damage feature. When scattering occurs from multiple centers, a coherent superposition of waves may occur, increasing the periodicity of the pattern to a fraction of the wavelength. We note the onset of new damage sites in between or an extension of the damage (see the case N = 20), supported by multiple scattering effects. The new surface sites have a dominant superficial character but a certain volume in fluence is seen with a slight modification of the material as seen experimentally in Fig.7(c), with a simulated sectional view given in Fig.7(d). The figures stresses the near-field effects for a focusing centered on the channel at low dose (N = 5) and the onset of a secondary site damage when focusing is displaced by the scan (N = 20) for about 600nm in a region likely experiencing incubation effects. Thus the scan will provide, by shifting the beam center, an excess of energy in a region that sufered prior excitation by field superposition.

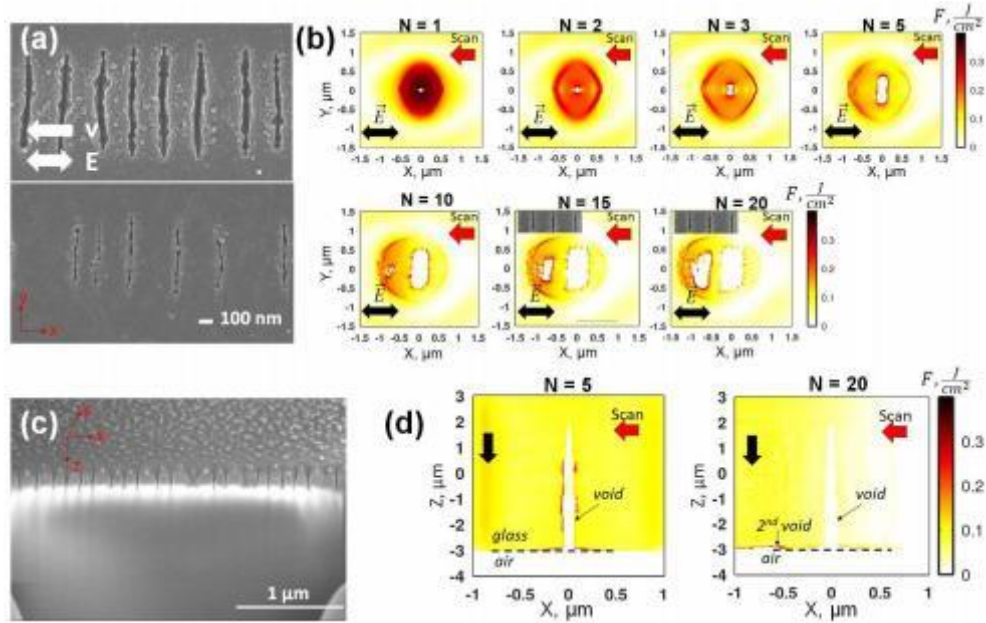

Figure 7: (a) Experimental evolution of quasi-periodic traces upon variation of input energy and pulse duration for a scan direction parallel to the laser pulse polarization (top 2ps at approximately 5μJ input energy, bottom 1ps at approximately 3μJ input energy). The period depends on the accumulated intensity dose. (b) Multipulse simulation of triggering new damage lines as a com-bination of near-field recovery and far-field scattering. Nonlinear fluence distributions F on the rear surface of glass correspond to different number of applied pulses N while scanning from right to left parallel to the field polarization starting from a tiny 100nm-nanovoid in the center of the grid [X,Y] = (0, 0). (c) Volume development of self-organized structures under light. (d) Sectional planes (X, Z) showing the simulated volumetric distribution of energy, with the enlargement of the void channel for N = 5 and with the onset of a secondary feature for N = 20.

The quasi-periodicity of the pattern raises the question of process similarity with regular laser-induced periodic surface structures (LIPSS) produced by scattered fields [26], and notably to the high frequency components. Regular LIPSS are developed by scattering surface waves from randomly scattering centers, creating a motif of constructive interference patterns, in-volving a range of near and far-field components [27] and pulse-to-pulse feedback, as recently discussed [28] in scanning conditions. A fundamental difference appears in the case of the struc-tures presented here. The process discussed [12, 25] relies on a deterministic process, seeded by a well-defined single scatterers and sequential line writing, with a degree of randomness in the position of the adjacent lines given by the threshold probability. This diverges from a regular LIPSS process triggered by a random and dense distribution of scatterers interacting together.

## 4 Conclusion

In conclusion, using far-field, loosely focused non-diffractive ultrafast infrared radiation with a signiicant working distance, we demonstrate the process of nanotrenching inside and on the backside of a glass sample with a remarkable section lying below $\lambda/100$ and, at the same time, a significant aspect ratio. The triggering element is the formation of a nanochannel with diameter of less than 100nm by direct far-field focusing and material nanocavitation that sets the process scale. Upon scanning the nanoscribing process is then driven by near-field components generated all along the longitudinal axis of the beam that etch the material. The process can be controlled by polarization on surface and in the volume. A 3D multipulse model coupling electromagnetic and material responses upon laser scanning describes the regimes of nanotrenching starting from a single nanovoid channel by near-field ablation and self-replication of multiple voids via far-fields depending on scan direction with respect to laser polarization. Far-field components also trigger the evolution of the process into self-organization under light. The results is not material specific and can have a larger action range. Such laser-based high aspect nanoprocessing capability at $\lambda/100$ opens interesting opportunities in nanophotonics, nanofluidics, and, in general, in designing metasurfaces and metamaterials, responding to a need of insertion of remote laser methods for nanoprocessing of a large range of materials.

## 5 Acknowledgment

National Key R&D Program of China(2022YFB4600200) and Natural Science Basic Research Program of Shaanxi Province (2022JQ-648) are gratefully acknowledged. RS has been partially supported by the French National Research Agency (ANR) with the grants ANR-19-CE30-0036 and ANR-21-CE08-0005.

## 6 Author contribution

GZ carried on the experiments, collected and analyzed the data. AR developed the model, carried out the modeling and provided the simulation data.GZ, RS, and GC designed the experiment, analyzed the experimental data, and wrote the manuscript. All authors discussed the data and revised and commented the manuscript.

## 7 Conflict of interest

The authors declare no conflict of interest.

## References

1. Stoian R. Ultrafast laser volume nanostructuring; a limitless perspective. Adv. Opt. Technol. 2023;12:1237524.
2. Ashkenasi D, Rosenfeld A, Varel H, W¨ahmer M, and Campbell E. Laser processing of sapphire with picosecond and sub-picosecond pulses. App. Surf. Sci. 1997;120:65– 80.
3. Garcia-Lechuga M, Ut eza O, Sanner N, and Grojo D. Evidencing the nonlinearity independence of resolution in femtosecond laser ablation. Opt. Lett. 2020;45:952–5.
4. Joglekar AP, Liu Hh, Meyh¨ofer E, Mourou G, and Hunt AJ. Optics at critical in- tensity: Applications to nanomorphing. Proc. Natl. Acad. Sci. USA 2004;101:5856– 61.
5. Schafer CB, Jamison AO, and Mazur E. Morphology of femtosecond laser-induced structural changes in bulk transparent materials. Appl. Phys. Lett. 2004;84:1441–3.

6. Englert L, Rethfeld B, Haag L, Wollenhaupt M, Sarpe-Tudoran C, and Baumert T. Control of ionization processes in high band gap materials via tailored femtosecond pulses. Opt. Express 2007;15:17855–62.

7. Korte F, Nolte S, Chichkov BN, et al. Far-ield and near-field material processing with. femtosecond laser pulses. Appl. Phys. A: Mat. Sci. Process. 1999;69:S7– S11.

8. Lei Y, Sakakura M, Wang L, et al. High speed ultrafast laser anisotropic nanostruc- turing by energy deposition control via near-field enhancement. Optica 2021;8:1365– 71.

9. Yan Z, Gao J, Beresna M, and Zhang J. Near-Field Mediated 40 nm In-Volume Glass Fabrication by Femtosecond Laser. Adv. Opt. Mater. 2022;10:2101676.

10. Li ZZ, Wang L, Fan H, et al. O-FIB: far-ield-induced near-field breakdown for direct nanowriting in an atmospheric environment. Light: Sci. Appl. 2020;9:41.

11. Li ZZ, Fan H, Wang L, et al. Super stealth dicing of transparent solids with nano- metric precision. Nat. Photonics 2024;18:799–808.

12. Lin Z, Liu H, Ji L, Lin W, and Hong M. Realization of 10nm Features on Semiconductor Surfaces via Femtosecond Laser Direct Patterning in Far Field and in Ambient Air. Nano Lett. 2020;20:4947–52.

13. Shimotsuma Y, Kazansky PG, Qiu J, and Hirao K. Self-Organized Nanogratings in Glass Irradiated by Ultrashort Light Pulses. Phys. Rev. Lett. 24 2003;91:247405.

14. Abou Saleh A, Rudenko A, Reynaud S, Pigeon F, Garrelie F, and Colombier JP. Sub-100 nm 2D nanopatterning on a large scale by ultrafast laser energy regulation. Nanoscale 12 2020;12:6609– 16.

15. Juodkazis S, Nishimura K, Tanaka S, et al. Laser-Induced Microexplosion Conined in the Bulk of a Sapphire Crystal: Evidence of Multimegabar Pressures. Phys. Rev. Lett. 16 2006;96:166101.

16. Bhuyan MK, Somayaji M, Mermillod-Blondin A, Bourquard F, Colombier JP, and Stoian R. Ultrafast laser nanostructuring in bulk silica, a 'slow' microexplosion. Optica 2017;4:951–8.

17. Bhuyan MK, Courvoisier F, Lacourt PA, et al. High aspect ratio nanochannel ma- chining using single shot femtosecond Bessel beams. Appl. Phys. Lett. 2010;97:081102.

18. Stoian R, Bhuyan MK, Rudenko A, Colombier JP, and Cheng G. High-resolution material structuring using ultrafast laser non-difractive beams. Adv. Phys. X 2019;4:1659180.

19. Rudenko A, Colombier JP, Itina TE, and Stoian R. Genesis of Nanogratings in Silica Bulk via Multipulse Interplay of Ultrafast Photo-Excitation and Hydrodynamics. Adv. Opt. Mater. 2021;9:2100973.

20. Grady DE. The spall strength of condensed matter. J. Mech. Phys. Solids 1988;36:353– 84.

21. Pacchioni G, Skuja L, and Griscom D. Defects in $SiO_2$ and Related Dielectrics: Science and Technology. 2000. doi: 10.1007/978-94-010-0944-7.

22. Nguyen HD, Tsaturyan A, Sao Joao S, et al. Quantitative Mapping of Transient Thermodynamic States in Ultrafast Laser Nanostructuring of Quartz. Ultrafast Sci. 2024;4:0056.

23. Rudenko A, Colombier JP, Höhm S, et al. Spontaneous periodic ordering on the surface and in the bulk of dielectrics irradiated by ultrafast laser: A shared electro- magnetic origin. Sci. Reports 2017;7:12306.

24. ktem B, Pavlov I, Ilday S, et al. Nonlinear laser lithography for indeinitely large- area nanostructuring with femtosecond pulses. Nat. Photonics 2013;7:897–901.

25. Liang F and Vallee R. Aperiodic nature of nanograting inscribed by femtosecond pulses. Opt. Express 2017;25:26124–32.

26. Sipe JE, Young JF, Preston JS, and Driel HM van. Laser-induced periodic surface structure. I. Theory. Phys. Rev. B 2 1983;27:1141–54.

27. Zhang H, Colombier JP, Li C, Faure N, Cheng G, and Stoian R. Coherence in ultrafast laser-induced periodic surface structures. Phys. Rev. B 17 2015;92:174109.

28. Sun J, Wang S, Zhu W, Li X, and Jiang L. Simulation of femtosecond laser-induced periodic surface

structures on fused silica by considering intrapulse and interpulse feedback. J. Appl. Phys. 2024;136:013103.